\documentstyle[12pt]{article}

\newcommand{\be}{\begin{equation}}
\newcommand{\ee}{\end{equation}}
\newcommand{\bea}{\begin{eqnarray}}
\newcommand{\eea}{\end{eqnarray}}

\renewcommand{\theequation}{\arabic{section}.\arabic{equation}}

\begin{document}
\begin{titlepage}

%\flushright{To Appear: }

\vspace{1in}

\begin{center}
\Large
{\bf Mirror Images of String Cosmologies}

\vspace{1in}

\normalsize

\large{James E. Lidsey}$^1$

\normalsize
\vspace{.7in}

{\em Astronomy Unit, School of Mathematical Sciences, \\
Queen Mary and Westfield College,\\  
Mile End Road, London, E1 4NS, U.K.}

\end{center}

\vspace{1in}

\baselineskip=24pt
\begin{abstract}
\noindent
A discrete symmetry of the four--dimensional string effective 
action is employed to derive
spatially homogeneous and 
inhomogeneous string cosmologies from 
vacuum solutions of general relativity that admit 
two commuting spacelike 
Killing vectors. In particular,  a tilted  
Bianchi type V cosmology is generated from a vacuum type 
${\rm VI}_h$ solution  and a plane wave solution with a bounded and 
oscillating dilaton field is found from a type ${\rm VII}_h$ 
model. Further applications are briefly discussed. 

\end{abstract}

PACS NUMBERS: 98.80, 04.50.+h, 11.25.Mj

\vspace{.7in}
$^1$Electronic mail: J.E.Lidsey@qmw.ac.uk
 
\end{titlepage}

\setcounter{equation}{0}

\def\theequation{\arabic{equation}}

The cosmological implications of string theory are currently 
receiving considerable attention 
\cite{pbb,gw,clw,b,ks}. Although a non--perturbative 
formulation of the theory has yet to be established, the low--energy 
effective action provides a framework for describing the evolution 
of the very early universe immediately below the string scale. 
This era represents a natural environment for establishing 
observational constraints on the theory and addressing 
central questions such as whether it admits  
realistic inflationary universes \cite{pbb,gw}. 

The search for cosmological solutions to the string equations of motion 
is therefore well motivated. The majority of studies to date 
have investigated spatially homogeneous backgrounds, including the 
isotropic Friedmann--Robertson--Walker (FRW) 
cosmologies \cite{clw}, the Bianchi models \cite{b}
and Kantowski--Sachs universes \cite{ks}. 
Since our understanding 
of the geometry of the universe near the string scale 
is incomplete, however, it is 
important to consider the role of inhomogeneities 
in string cosmology. 
The field equations for inhomogeneous backgrounds are 
in general very difficult to solve, but further 
progress can be made by considering models where  
homogeneity is broken along one spatial direction. String 
cosmologies satisfying this property were recently 
derived by employing a variety of methods \cite{inh}. 

The purpose of the present letter is to show that a 
wide class of homogeneous and 
inhomogeneous string cosmologies may be generated in a 
very straightforward manner from known vacuum solutions 
of general relativity. We employ a discrete ${\rm Z}_2$ symmetry  
that arises in the Neveu--Schwarz/Neveu--Schwarz (NS--NS) 
sector of the effective action  when the space--time 
admits two commuting spacelike Killing vectors \cite{bakas}. 

To lowest order in the inverse string tension, 
$\alpha'$, the four--dimensional effective NS--NS action is \cite{perry}
\be
\label{effective}
S=\int d^4 x \sqrt{-G} e^{-\Phi} \left[ R (G)+ \left( \nabla \Phi \right)^2 
-\frac{1}{12} H_{\mu\nu\lambda}H^{\mu\nu\lambda} \right]   ,
\ee
where $R$ is the Ricci curvature scalar of the space--time with 
metric $G_{\mu\nu}$, 
$\Phi$ is the dilaton field, $H_{\mu\nu\lambda} \equiv 
\partial_{[\mu} B_{\nu\lambda ]}$ is the 
field strength of the antisymmetric two--form potential, 
$B_{\mu\nu}$, and $G \equiv {\rm det} G_{\mu\nu}$. 
It is convenient to perform the conformal 
transformation to the `Einstein frame':
\be
\label{conformal}
g_{\mu\nu} \equiv \Theta^2 G_{\mu\nu} , \qquad \Theta^2 \equiv e^{-\Phi}  .
\ee
The effective action (\ref{effective}) then takes the form   
\be
\label{effact}
S=\int d^4 x \sqrt{-g} \left[ R (g) -\frac{1}{2} \left( 
\nabla \Phi \right)^2 -\frac{1}{12} e^{-2\Phi} 
H_{\mu\nu\lambda} H^{\mu\nu\lambda} \right]  ,
\ee
where $g \equiv {\rm det} g_{\mu\nu}$. 

In four--dimensions, the field strength of the two--form potential is dual 
to a one--form. We may therefore apply a duality transformation:
\be
H^{\mu\nu\lambda} \equiv \epsilon^{\mu\nu\lambda\kappa} e^{2\Phi} 
\nabla_{\kappa} \sigma  ,
\ee
where $\sigma$ represents a pseudo--scalar axion field and 
$\epsilon^{\mu\nu\lambda\kappa}$ is the covariantly constant 
four--form. The action (\ref{effact}) is then equivalent to 
the non--linear sigma--model
\be
\label{eineff}
S= \int d^4 x \sqrt{-g} \left[ R +\frac{1}{4} {\rm Tr} 
\left( \nabla N \nabla N^{-1} \right) \right]    ,
\ee
where
\be
N \equiv \left( \begin{array}{cc}
e^{\Phi} & \sigma e^{\Phi} \\
\sigma e^{\Phi} & e^{-\Phi} + \sigma^2 e^{\Phi} \end{array} 
\right) 
\ee
is a symmetric ${\rm SL}(2, R)$ matrix. 
The dilaton and axion parametrize the ${\rm SL}(2, R)/{\rm 
U}(1)$ coset and Eq. (\ref{eineff}) is invariant 
under global ${\rm SL}(2,R)$ transformations: 
\be
\label{sduality}
\tilde{g}_{\mu\nu} = g_{\mu\nu} , \qquad \tilde{N} = \Omega N 
\Omega^T    ,
\ee
where 
\be
\Omega  \equiv \left( \begin{array}{cc}
d & c \\
b & a \end{array} \right) , \qquad ad -bc =1
\ee
is a constant ${\rm SL}(2,R)$ matrix. 

We consider the class of metrics defined by the block diagonal line element 
\be 
\label{9}
ds^2 =h_{\alpha\beta} (x^{\epsilon}) dx^{\alpha}dx^{\beta} 
+ \gamma_{ab} (x^{\epsilon}) dx^a dx^b    ,
\ee
where $\{ x^{\alpha} =t,x \}$ and $\{ x^a =y,z \}$. The functions 
$h_{\alpha\beta}$ and $\gamma_{ab}$ depend only on the variables 
$t$ and $x$ and spatial homogeneity is broken along the $x$--direction
\cite{review,review1}. Eq. 
(\ref{9}) admits two commuting 
spacelike Killing vectors, $\partial /\partial x^a$, and there exists 
an abelian group, $G_2$, of isometries. We refer to this class of 
metrics as $G_2$ models. 

If all  
massless excitations in the string action (\ref{eineff}) 
are functions only of $t$ 
and $x$, we may integrate over 
the transverse coordinates to derive an effective two--dimensional 
action:
\be
\label{2d}
S=\int d^2 x \sqrt{-h} e^{-\gamma} \left[ R_2 
+\frac{1}{2} \left( \nabla \gamma \right)^2 + 
\frac{1}{4} {\rm Tr} \left( \nabla f \nabla f^{-1} \right) 
+\frac{1}{4} {\rm Tr} \left( \nabla N \nabla N^{-1} 
\right) \right]   ,
\ee
where $R_2$ is the Ricci curvature of the $(1+1)$--dimensional manifold 
with metric $h_{\alpha\beta}$, $h \equiv {\rm det}h_{\alpha\beta} $ and 
$\gamma \equiv - (\ln {\rm det} \gamma_{ab} )/2$. The 
symmetric $2 \times 2$ 
matrix, $ f_{ab} 
\equiv e^{\gamma} \gamma_{ab}$, is an element of ${\rm SL}(2,R)$ 
and may be expressed in the form  
\be
\label{f}
f \equiv \left( \begin{array}{cc}
e^V & \omega  e^V \\
\omega e^V & e^{-V} +\omega^2 e^V \end{array}
\right)   .
\ee
Thus, the scalar functions $V$ and $\omega$ parametrize a second 
${\rm SL}(2, R)/{\rm U}(1)$ coset. 

The field equations derived from Eq. 
(\ref{2d}) admit  an infinite--dimensional 
symmetry that can be identified infinitesimally 
with  the ${\rm O}(2,2)$ current algebra \cite{bakas,mah1,kehagias}. 
The well--known non--compact, global  ${\rm O}(2,2)$ and ${\rm SL}(2, R)$ 
symmetries may be embedded in this larger symmetry. 
As discussed by Bakas \cite{bakas}, 
however, there exists a further ${\rm Z}_2$ 
symmetry that corresponds to a discrete interchange of the 
field content of the two 
${\rm SL}(2, R)/{\rm U}(1)$ cosets in Eq. (\ref{2d}): 
\be
\label{d}
\tilde{N}_{ab} = f_{ab} , \qquad \tilde{f}_{ab} = N_{ab}   .
\ee
The two--metric, $h_{\alpha\beta}$, and determinant, $\gamma$, 
are invariant under this transformation. 

Eq. (\ref{d}) is not 
part of the ${\rm SL}(2,R)$  transformation (\ref{sduality}) because the 
Einstein--frame metric, $g_{\mu\nu}$, does not transform as a singlet. 
Moreover, it is not part of the global ${\rm O}(2,2)$
symmetry \cite{mah}. This symmetry leaves 
the shifted dilaton field, $\phi \equiv \Phi + \gamma$, 
invariant and, in 
general, the dilaton is not a singlet under Eq. (\ref{d}), whereas 
$\gamma$ is. 
In effect, Eq. (\ref{d}) interchanges the dilaton and 
axion fields with the components of the metric 
on the surfaces of orthogonality. 
The axion field is interchanged 
with the off--diagonal component, $\omega$, in Eq. (\ref{f}) 
and the 
dilaton field with the function $V$. 

A string background is parametrized in terms of the 
massless degrees of freedom $\{ G_{\mu\nu} , \Phi , B_{\mu\nu} \}$. 
In this sense, a vacuum solution to Einstein gravity may be viewed 
as the subset of string backgrounds, 
$\{ G_{\mu\nu} , 0, 0 \}$, where the dilaton and two--form potential 
are trivial. 
Application of Eq. (\ref{d}) to a  given string background then   
leads to a new solution $\{ \tilde{G}_{\mu\nu} , 
\tilde{\Phi} , \tilde{B}_{\mu\nu} \}$ that admits a 
different space--time interpretation. The latter solution 
may be viewed as the `mirror image' of the former. 

We now employ the invariance (\ref{d}) to derive inhomogeneous string 
cosmologies containing both dilaton and axion fields from vacuum 
$G_2$ solutions. The Gowdy models are the class of vacuum $G_2$ cosmologies 
where the spacelike hypersurfaces are compact \cite{gowdy}.  
The allowed topologies are a three--torus, $S^1 \times S^1 \times S^1$, 
a hypertorus, $S^1 \times S^2$, and a three--sphere, $S^3$. Without 
loss of generality,  the toroidal models may be written in the form 
\be
\label{gow}
ds^2 =e^k \left( -dt^2 +dx^2 \right) + t \left( e^p dy^2 
+ e^{-p} dz^2 \right)   ,
\ee
where $k$ and $p$  represent the longitudinal and transverse 
parts of the gravitational field, respectively, 
and are functions of $t$ and $x$. The vacuum Einstein field 
equations then reduce to 
\bea
\label{field1}
\ddot{p} +\frac{1}{t} \dot{p} - p'' =0 \\
\label{field2}
\dot{k} =-\frac{1}{2t} +\frac{t}{2} \left( \dot{p}^2 + {p'}^2 \right) \\
\label{field3}
k' =t\dot{p} p'   ,
\eea
where a dot and prime denote $\partial /\partial t$ and 
$\partial / \partial x$, respectively. The general 
solution to Eqs. (\ref{field1})--(\ref{field3}) consistent with 
the toroidal boundary conditions is known \cite{gowdy,cm}. 

Since the metric (\ref{gow}) is diagonal, a direct application 
of Eq. (\ref{d}) would generate a string solution 
with only the dilaton field excited. The simplest 
method for introducing an off--diagonal term  in the metric 
is to perform an ${\rm SL}(2, R)$ transformation  in the two--space 
of the Killing vectors,  $\partial / \partial x^a$: 
\be
\label{Ehlers}
\tilde{f} = \Theta f \Theta^T , \qquad 
\Theta \equiv \left( \begin{array}{cc}
D & C \\
B & A \end{array} \right)   ,
\ee
where $AD -BC =1$ and all other variables are invariant 
\cite{ehlers}. The transverse metric (\ref{f}) 
then transforms to  
\bea
\label{Eh1}
e^{\tilde{V}} = C^2 e^{-V} +D^2 e^{V} \\
\tilde{\omega} = \frac{AC e^{-V} + BD e^{V}}{C^2 e^{-V} +D^2 e^{V}}  .
\eea

Eq. (\ref{Ehlers}) does not commute with the discrete 
transformation (\ref{d}) and the two may be employed together 
to generate the axion field. 
Applying Eq. (\ref{Ehlers}), 
followed by the discrete transformation (\ref{d}), 
leads to a new class of 
inhomogeneous string cosmologies with 
non--trivial dilaton and axion fields: 
\bea
\label{mirrormetric}
ds^2 = e^{k} \left( -dt^2 +dx^2 \right) + t \left( dy^2 + 
dz^2 \right) \\
\label{mirrordilaton}
\Phi = \ln \left[ C^2 e^{-p} +D^2 e^{p} \right] \\
\label{mirroraxion}
\sigma = 
\frac{AC e^{-p} + BD e^{p}}{C^2 e^{-p} +D^2 e^{p}}   ,
\eea
where $p$ solves Eq. (\ref{field1}). The metric in the string frame is then 
given by substituting Eqs. (\ref{mirrormetric}) and (\ref{mirrordilaton}) 
into Eq. (\ref{conformal}): 
\be
\label{mirrorstring}
dS^2 =e^{k+\Phi} \left( -dt^2 +dx^2 \right) + te^{\Phi} \left( 
dy^2 +dz^2 \right)    .
\ee

There is no preferred direction in the 
transverse space in Eq. (\ref{mirrormetric}) and this implies the existence 
of a one--parameter isotropy group in addition to the $G_2$ isometry 
group. 
The mirror images of the toroidal Gowdy models 
are therefore locally rotationally symmetric (LRS) 
inhomogeneous cosmologies with plane symmetry. We remark that 
a similar analysis may be applied to the hypertorus and three--sphere 
Gowdy models. 

We now discuss further  applications of Eq. (\ref{d}) 
within the context of the 
spatially homogeneous cosmologies that admit  
an abelian subgroup $G_2$ of isometries. This includes the Bianchi types 
I--VII and the LRS types VIII and IX \cite{lrs}. 
The general spatially homogeneous vacuum solution  
with a simply transitive Lie group $G_3 = \Re^3$ is 
the type I Kasner solution \cite{kasner}: 
\be
\label{Kasner} 
ds^2 =-dt^2 +t^{2p_1} dx^2 + t^{2p_2} dy^2 + t^{2p_3} dz^2    , 
\ee
where $\sum_{i=1}^3 p_i =\sum_{i=1}^3 p_i^2 =1$. 
Applying Eqs. (\ref{Ehlers}) and Eq. (\ref{d}) yields the LRS type I solution: 
\bea
\label{mirrorI}
ds^2 = -dt^2 +t^{2p_1}dx^2 +t^{p_2+p_3} \left( dy^2 + 
dz^2 \right) \nonumber \\
\Phi = \ln \left[ C^2 t^{p_3 -p_2} +D^2 t^{p_2 -p_3} \right]
\nonumber \\
\sigma = \frac{ACt^{p_3-p_2} +BDt^{p_2-p_3}}{C^2t^{p_3-p_2} 
+D^2t^{p_2-p_3}}   
\eea
and this reduces to the general, spatially flat FRW string cosmology
when $p_2 = (1 -\sqrt{3})/3$ and $p_3 =
(1+\sqrt{3})/3$ \cite{clw}. 

One of the simplest examples of a spatially curved, anisotropic model 
is the Bianchi type V. A vacuum type V solution was found by Joseph 
\cite{joseph}:
\be
ds^2 = \sinh 2t \left( -dt^2 +dx^2  
+e^{-2x} \left[ \left( {\rm tanh} t \right)^{\sqrt{3}} 
dy^2 + \left( {\rm tanh} t \right)^{-\sqrt{3}} dz^2 
\right] \right)  
\ee
and Eqs. (\ref{Ehlers}) and (\ref{d}) yield its mirror image: 
\bea
\label{FRW-1}
ds^2 = \sinh 2t \left[ -dt^2 +dx^2 +e^{-2x}\left( dy^2 +dz^2 \right) 
\right] \nonumber \\
\Phi = \ln \left[ C^2 \left( {\rm tanh} t \right)^{-\sqrt{3}} 
+ D^2 \left( {\rm tanh} t \right)^{\sqrt{3}} \right]  \nonumber \\
\sigma = \frac{AC \left( {\rm tanh} t \right)^{-\sqrt{3}} +BD 
\left( {\rm tanh} t \right)^{\sqrt{3}}}{C^2 \left( 
{\rm tanh} t \right)^{-\sqrt{3}} +D^2 \left( 
{\rm tanh} t \right)^{\sqrt{3}}}    .
\eea
This is the general solution for the  negatively curved FRW string 
cosmology \cite{clw}. 

The Joseph vacuum solution was generalized by Ellis and MacCallum to 
Bianchi type ${\rm VI}_h$ \cite{em}: 
\be
ds^2 = \sinh 2t \left[ A^b \left( -dt^2 +dx^2 \right)
+A e^{2(1+b)x} dy^2 +A^{-1} e^{2(1-b)x} dz^2 \right]   ,
\ee
where 
\be
A \equiv \left( \sinh 2t \right)^b  \left( {\rm tanh} t 
\right)^{\sqrt{3+b^2}}
\ee
and $b^2 \equiv -1/h$. This corresponds to the 
Joseph type V solution when $b=0$. 
The mirror image of the Ellis--MacCallum type  ${\rm VI}_h$ cosmology is
\be
\label{mirror6}
ds^2 =\sinh 2t \left[ A^b \left( -dt^2 +dx^2 \right) +e^{2x} \left( dy^2 
+dz^2 \right) \right]  ,
\ee
where the dilaton and axion fields are given by Eqs. (\ref{mirrordilaton}) 
and (\ref{mirroraxion}), respectively, with  
\be
\label{*}
e^p =\left( \sinh 2t \right)^b \left( {\rm tanh} t \right)^{\sqrt{3+b^2}} 
e^{2bx}   .
\ee

A calculation of the structure 
constants of the isometry group of 
Eq. (\ref{mirror6}) implies that it is a LRS Bianchi type V 
cosmology. Furthermore, the fluid flows associated 
with the dilaton and axion fields are not 
orthogonal to the surfaces of homogeneity, 
$t={\rm constant}$. Thus, Eq. (\ref{mirror6}) represents a tilted 
LRS Bianchi type V string cosmology. 

Finally, we consider the vacuum type ${\rm VII}_h$ plane wave 
solution \cite{dlv}:
\be
\label{lukash}
ds^2 =-dt^2 + s^{-2} t^2 dx^2 + t^{2s} e^{-2x} dl^2   ,
\ee
where 
\be
dl^2 = \cosh \mu  \left( dy^2 +dz^2 \right) -\sinh \mu \left[ 
\left( dy^2 -dz^2  \right) \cos 2ku +2 dy dz \sin 2 k u \right]
\ee
and $ u \equiv x -s \ln t$, $s^{-1} \equiv 1+ k^2 {\sinh}^2 
\mu$ and $\{ k , \mu \}$ are constants. This metric may be 
interpreted as two monochromatic, circularly--polarized 
gravitational waves with vectors $\pm {\rm {\bf k}}$ travelling 
in the $\pm x$ directions with constant amplitudes $\mu$
\cite{lukash}. It 
can be shown that this metric admits a covariantly constant null 
Killing vector  field, $l_{\nu} \equiv \partial_{\nu}u$, 
$l_{\nu}l^{\nu} =0$, and is therefore 
an exact solution to the classical string equations of motion 
to {\em all} orders in the inverse string tension 
\cite{hs}. 

The mirror image of the Lukash solution is
\be
\label{osc}
ds^2 = -dUdV  + U^{2s} \left( dy^2 +dz^2 \right)   ,
\ee
where $U\equiv te^{-x/s}$ and $V \equiv te^{x/s}$ and  
the dilaton and axion fields are given by
\bea
\label{7d}
e^{\Phi} = \cosh \mu -\sinh \mu \cos 2ku \\
\sigma = - \frac{\sin 2ku \sinh \mu}{\cosh \mu -\sinh \mu \cos 2ku}  ,
\eea
respectively. The energy momentum tensor 
for the matter fields may be written in the form $T_{\mu\nu} 
=\Pi (u) l_{\mu} l_{\nu}$, where $\Pi =\Pi (u)$ is a scalar function of the 
light--cone coordinate $u$. Thus, 
Eq. (\ref{osc}) may be interpreted as a conformally flat
plane wave background with pure 
radiation (null dust). 
It is interesting because it is also exact to all order in $\alpha'$ 
due to the existence of the covariantly constant null Killing vector field. 
The dilaton field oscillates with a constant frequency and 
is bounded from above and below. 
In the string frame, this field parametrizes the 
effective gravitational (string) 
coupling, $g_s^2 \equiv e^{\Phi}$,  and Eq. (\ref{osc}) is therefore 
an example of an exact string solution with an oscillating, but 
bounded,  gravitational `constant'.  

We conclude with a discussion of further applications 
of the symmetry (\ref{d}). It is important 
to emphasize that different space--time interpretations 
apply to solutions related by Eq. (\ref{d}). This is evident 
from  the simple derivation of the general, negatively curved FRW 
string cosmology from the Jacobs vacuum type V solution. 
In general, Eq. (\ref{d})  may be 
employed to derive inhomogeneous $G_2$ string cosmologies 
with non--trivial dilaton and axion fields from 
solutions where neither  of these fields is 
initially trivial. This 
includes backgrounds 
constructed from conformal field theories (see Ref. \cite{bakas} and 
references therein). 

These string models have a number of important physical applications.
They allow density perturbations 
in string--inspired inflationary models such as the pre--big bang 
scenario to be studied \cite{pbbh}. The  
propagation of gravitational waves in string backgrounds 
may also be analyzed in terms of $G_2$ space--times \cite{adams}.
Furthermore, the collision of self--gravitating plane waves can be 
modelled as the time reversal  
of a $G_2$ cosmology in the vicinity of the big bang 
singularity \cite{gri}. 

We applied the discrete symmetry (\ref{d}) within the context of vacuum 
general relativity. This is interesting because exact, 
vacuum $G_2$ solutions to Einstein 
gravity have been extensively studied \cite{Mac}
and a large number of solution--generating techniques 
are known \cite{review1}. New  
algorithms that include Eq. (\ref{d}) may be developed 
to generate the corresponding 
string cosmologies. For example, there exist 
a number of methods for generating inequivalent $G_2$ vacuum 
solutions with off--diagonal terms in the 
spatial metric \cite{scatter}. Together with Eq. (\ref{d}), these  
will lead to different string solutions  
to those derived in this work. Alternatively, 
cosmological models with a single, minimally 
coupled scalar field may be generated from vacuum 
solutions by means of the algorithm introduced by Barrow \cite{jb} and 
developed by Wainwright, Marshman and Ince 
\cite{wmi}. These solutions could then serve as seeds 
for generating string backgrounds in the manner outlined above. 
Moreover, once a solution with non--trivial dilaton and two--form 
potential has been derived in this way, 
more general type II string backgrounds with 
non--trivial Ramond--Ramond fields can  be found directly \cite{RR}. 

Finally, we outline a related method for deriving 
four--dimensional string cosmologies from vacuum 
Einstein gravity. The ${\rm SL}(2, R)$ symmetry (\ref{sduality}) 
of the effective action (\ref{eineff}) may be interpreted geometrically 
by considering the compactification 
of six--dimensional vacuum Einstein gravity,  
$S=\int d^6x \sqrt{-g_6} R_6$, on a non--dynamical two--torus. 
Assuming the {\em ansatz}
\be
\label{6}
ds_6^2 = {^{(4)}}g_{\mu\nu} (x) dx^{\mu}dx^{\nu} 
+ e^{-\Phi (x)}dy_6^2 +e^{\Phi (x)} \left( dy_5 + \sigma (x) dy_6 
\right)^2
\ee
for the higher--dimensional metric and  
integrating over the spatial variables $y^i$ $(i=5,6 )$ then 
leads to an effective four--dimensional 
action that is formally equivalent to 
Eq. (\ref{eineff}) if the dilaton and axion fields are identified with 
the appropriate degrees of freedom in the metric (\ref{6}). 

Thus, four--dimensional string backgrounds may be derived directly 
from six--dimensional, Ricci--flat solutions. 
This geometrical interpretation is 
similar to the recently proposed conjecture 
that the ${\rm SL}(2, Z)$ symmetry of the ten--dimensional
type IIB superstring arises from the toroidal compactification of 
twelve--dimensional F--theory \cite{F}. 
It is closely related to 
a theorem of differential geometry due to Campbell that states that  
a Riemannian manifold of dimension $n$ may be locally 
and isometrically embedded in a Ricci--flat, Riemannian manifold 
of dimension $N\ge n+1$ \cite{campbell}. 
It would be interesting to investigate this correspondence 
further to derive non--trivial string models. 
Applying Campbell's theorem to embed lower--dimensional manifolds in 
six--dimensions would lead to a 
wide class of Ricci--flat solutions
of the form given by Eq. 
(\ref{6}). The $\{ y^i \}$
components of the embedding six--dimensional metric would then 
directly determine 
the behaviour of the dilaton and axion fields in four dimensions.

\vspace{.3in}
\centerline{\bf Acknowledgments}
\vspace{.3in}
The author is supported by the Particle Physics and Astronomy 
Research Council (PPARC), UK. We thank J. D. 
Barrow,  A. Coley, K. E. Kunze 
and D. Wands for helpful discussions.

\end{document}